\documentclass[twocolumn]{aastex631}

\usepackage{graphicx}
\usepackage{soul}
\usepackage{savesym}
\usepackage{cancel}
\savesymbol{tablenum}
\usepackage{siunitx}
\restoresymbol{SIX}{tablenum}
\usepackage{amsmath}
\usepackage{natbib}

\usepackage{hyperref}
\usepackage{orcidlink}

\newcommand{\orcid}[1]{\orcidlink{#1}}

\begin{document}

%% Reintroduced the \received and \accepted commands from AASTeX v5.2
\received{2024 December 4}
\revised{2025 May 13}
\accepted{2025 May 20}
%% Command to document which AAS Journal the manuscript was submitted to.
%% Adds "Submitted to " the argument.
% \submitjournal{ApJ}

\submitjournal{AAS}

\author{Pouya Tanouri\orcid{0009-0006-3261-6397}}
\affiliation{Department of Physics and Astronomy, University of British Columbia, Vancouver, BC, V6T1Z1, Canada}
\author{Ryley Hill\orcid{0009-0008-8718-0644}}
\affiliation{Department of Physics and Astronomy, University of British Columbia, Vancouver, BC, V6T1Z1, Canada}
\author{Douglas Scott\orcid{0000-0002-6878-9840}}
\affiliation{Department of Physics and Astronomy, University of British Columbia, Vancouver, BC, V6T1Z1, Canada}
\author{Edward L. Chapin\orcid{0009-0003-8033-1351}}
\affiliation{National Research Council of Canada, Herzberg Astronomy and Astrophysics Research Centre, Victoria, BC, V9E 2E7, Canada}
\shortauthors{Tanouri, Hill, Scott \& Chapin}

\correspondingauthor{Pouya Tanouri}
\email{ptanouri@phas.ubc.ca}

\title{\textbf{Improving Optical Photo-\textit{z} Constraints for Dusty Star-forming Galaxies Using Submillimeter-Based Priors}}
\shorttitle{Optical plus submm photo-\textit{z}s}

\begin{abstract}

Photometric redshifts (photo-\textit{z}s) provide an efficient alternative to spectroscopic redshifts, enabling redshift estimation for large galaxy samples. However, traditional photo-$z$ methods primarily rely on optical and near-infrared (OIR) photometry, which can struggle with dusty star-forming galaxies (DSFGs) that are often faint in the OIR but bright at far-infrared (FIR) and millimeter wavelengths. We present a method for incorporating FIR-to-millimeter photometry as a prior within standard OIR-based photo-$z$ frameworks, explicitly folding in the observed empirical relationship between total infrared luminosity and dust temperature. This approach is particularly suitable for wide-area surveys, such as those anticipated with the \textit{Euclid} satellite or Rubin Observatory, where OIR photo-$z$s can be complemented with longer wavelength data to help with the dustiest and highest star-forming galaxies. Applying this method to the H-ATLAS catalog, which combines FIR photometry from {\it Herschel/}-SPIRE with OIR observations, we achieve a threefold reduction in catastrophic outliers compared to traditional OIR-based photo-\textit{z} techniques, demonstrating its utility for improving redshift estimates in FIR-bright galaxies.
% \setstcolor{red}

% \st{Some overstruck text}
\end{abstract}

\keywords{galaxies: photometry, far-infrared ---  galaxies: dusty, starbursts  --- submillimeter: galaxies}

\section{Introduction} \label{sec:intro}

Accurate redshift estimation is essential for understanding the intrinsic properties of galaxies and using them as cosmological probes. Redshifts are determined by identifying features in a galaxy’s spectral energy distribution (SED), such as emission lines, absorption lines, and breaks, which shift to longer wavelengths due to the expansion of the Universe \citep[e.g.,][]{Baum1957,Yee1998,NewmanGruen2022}.

While spectroscopy provides the most accurate redshifts (henceforth spec-\textit{z}s) thanks to its high spectral resolution, photometric redshifts (henceforth photo-\textit{z}s) are much more efficient and are therefore the primary tool for large galaxy surveys. This is especially true for fainter sources, since spectroscopic redshifts require deeper integrations than photometric redshifts and so are not always readily available due to constraints on telescope time, or are only available for a biased subset of brighter sources (see e.g., \citealt{salvato2018flavours}). Despite advancements in multi-object spectrographs over the past decade, we are only able to acquire adequate spectra for a small percentage of sources detected in deep imaging surveys.

Most photo-\textit{z} methods limit themselves to only optical and near-infrared (NIR) data. In these wavebands, the main broadband SED signatures are the 4000\,\AA{} Balmer break, arising from the absorption of photons with energies above the Balmer limit and the superposition of various ionized metal absorption lines in stellar atmospheres, and the Lyman break (below 1216\,\AA{}), caused by the absorption of photons with energies beyond the Lyman limit \citep{ Ilbert_2008}. 

However, for dusty star-forming galaxies (DSFGs), while emission lines can be intrinsically strong due to high SFRs, significant dust attenuation often obscures these features in the optical and near-infrared (OIR).
Yet there is an additional broadband feature accessible at longer far-infrared (FIR) and submillimeter (submm) wavelengths, resulting from the thermal emission of dust heated by stars. The use of this feature may allow us to improve photo-\textit{z}s for these sources, particularly in cases where OIR photometry is limited in depth or affected by significant dust attenuation. DSFGs are clearly an important class of galaxy, responsible for roughly half of the cosmic star-formation rate (SFR) at intermediate redshifts \citep{koprowski2017}, and so obtaining reliable redshifts for this population is crucial for furthering galaxy evolution.

Figure~\ref{fig:EAZY_outliers} illustrates an example of photo-\textit{z}s and spec-\textit{z}s from the Great Observatories Origins Deep Survey (GOODS) of the Chandra Deep Field-South (CDF-S), color coded by $B-V$ color. A notable discrepancy between the photo-\textit{z}s and spec-\textit{z}s is observed for a small fraction (about 5\%) of galaxies, { with some prefernce for the redder galaxies. Such discrepancies are often referred to as `catastrophic outliers'. These galaxies tend to be redder in all bands \citep{Wuyts2008}, meaning that they are often DSFGs \citep[e.g.][]{Dokkum2011}. This discrepancy could be attributed to the degeneracy caused by misidentifying the Balmer break with the Lyman break (or vice versa) at different redshifts, or due to the lack of any prominent break signatures in the SED at all. %DSFGs and AGN can be bright at longer wavelengths. 
By incorporating additional information at FIR and submm wavelengths that track dust emission, we aim to resolve some of these catastrophic outliers. %These additional data provide a crude (but helpful) redshift estimate that can assist in choosing between peaks in the photo-\textit{z} likelihood distribution.
% citing the earlier Carilli & Yun
% photo-z paper which used the submm-radio color
The use of submm photometry to estimate DSFG redshifts has certainly been discussed before \citep[e.g.,][]{Carilli_1999, Blain2002, Aretxaga2003, Chakrabarti2013, SCUBA2}. 
A similar approach was taken by \citet{casey2020}, where the MMPZ algorithm utilized the observed distribution of galaxy SEDs ($0 < z < 5$) to derive photometric redshifts by linking the rest-frame peak wavelength and total IR luminosity through an empirical relationship. \citet{Manning_2022} further applied the MMPZ method, combining it with OIR photo-$z$s in a similar way to refine FIR/mm photometric redshifts.

However, here we explicitly present the calculation of how to improve photo-\textit{z} constraints by setting a prior based on submm data, and aim to fold independent analyses of optical and NIR SED photometry with FIR/submm photometry in a self-consistent manner, avoiding the assumptions required by energy balance calculations as employed by some photometric redshift-fitting codes (e.g., \texttt{CIGALE}, \citealt{boquien2019}). While energy balance can be effective in some cases, it may not fully account for dust self-absorption in optically thick regions, as seen in local ultraluminous infrared galaxies (ULIRGs), making it less suitable for DSFGs with significant attenuation. As pointed out by \citet{Nayyeri_2017}, this approach might be especially beneficial for forthcoming large multi-band extragalactic surveys such as the Cerro Chajnantor Atacama Telescope \citep[CCAT;][]{ccat2023} in the submillimeter range, and \textit{Euclid} \citep{Mellier2024} or Rubin \citep{rubin2019} in the optical/NIR.

This paper is organised as follows.  Section~\ref{sec:FIR_photometry} describes the mathematical framework for estimating photo-$z$s from FIR/submm photometry, Section~\ref{sec:optical} describes how we use a generic SED-fitting code to compute photo-\textit{z}s from optical and NIR photometry and combine it with our FIR/submm photo-$z$s, and in Section~\ref{sec:combining} we apply our procedure to real galaxy catalogs. We conclude with a summary of our findings and their implications in Section~\ref{sec:conclusions}.

\begin{figure}[htbp!]
\centering
  \includegraphics[scale = 0.35]{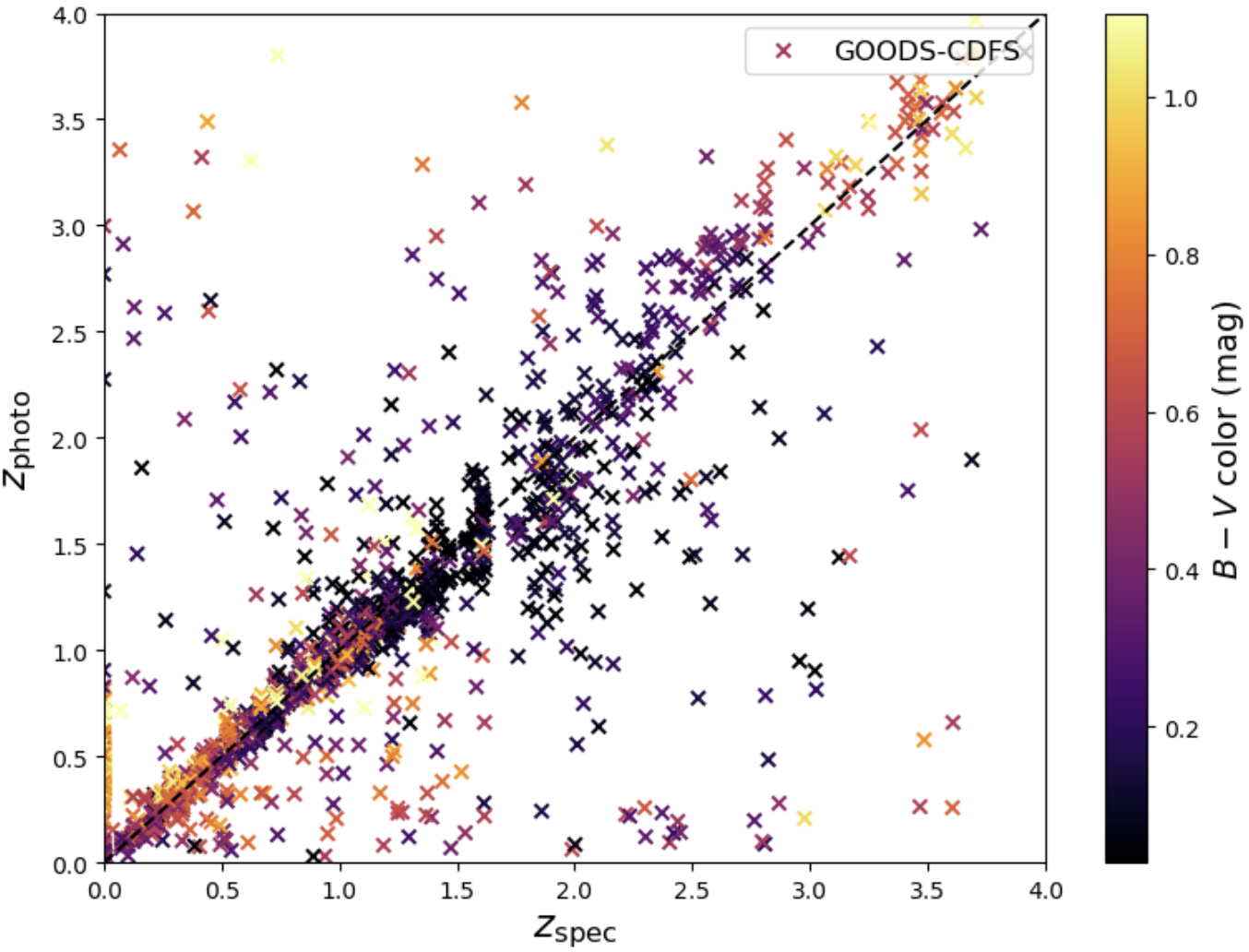}
  \caption{Comparison of spectroscopic and photometric redshifts using deep optical and NIR photometry from the GOODS CDF-S survey \citep{Wuyts2008}. 
  % Here $\sigma$ is defined as the commonly used metrics: the normalized median absolute deviation of the residuals given by $\sigma = 1.4826 \times \text{median}(|\Delta z - \text{median}(\Delta z)|)$,  
  Here the outlier fraction is defined as $ |z_{\text{spec}}-z_{\text{photo}}|/(1+z_{\text{spec}}) > 3\sigma$ (see Section~\ref{sec:combining} for more details), which is about 4\%. We expect many of these outliers to be DSFGs, and more accurate photometric redshifts can be obtained by incorporating FIR and submm photometry.  Labeling the galaxies with color ($B-V$ here) shows a preference for redder galaxies to be outliers.}
  % is the scaled residual between the reference and the photometric redshifts.}
  \label{fig:EAZY_outliers} 
\end{figure}

\section{Photometric redshift fitting in FIR/submm bands}\label{sec:FIR_photometry}
The process of star formation involves dust, which is subsequently heated by nearby stars to produce a thermal SED peaking at around $100\,\mu$m in the rest frame; consequently, DSFGs are bright at FIR/submm wavelengths as this emission becomes redshifted \citep{Franceschini1991, Blain_Longair1993}. 
{\it Herschel\/}-SPIRE \citep{Griffin2010} operated in three FIR bands centered on 250, 350, and 500\,$\mu\text{m}$. It conducted wide surveys of the sky, providing SED information for tens of thousands of DSFGs, including those that are faint in the optical and NIR. Here we use SPIRE photometry as an example to describe how to constrain photometric redshifts.  However, we note that our method can be generalised to photometry at arbitrary FIR/submm wavelengths. The code to do the fitting is publicly available.\footnote{https://github.com/ptanouri/Optical-FIR-Photo-z/tree/main}

\subsection{Mathematical background}

Considering that the rest-frame FIR emission from DSFGs is often approximated as being dominated by optically-thin dust, particularly in the Rayleigh-Jeans tail \citep[as reviewed by][]{Casey_2014}, we can model the emitted flux density in the rest-frame across a bandwidth $d\nu_{\textrm{em}}$ as a modified blackbody function:

\begin{equation}
\label{emitted_modified_bb}
     S_{\nu_\textrm{em}} ( T_{\textrm{em}}, A_{\textrm{em}}) d\nu_\textrm{em}= A_{\textrm{em}}\frac{\nu_{\textrm{em}}^{3+\beta}}{\text{exp}(h\nu_{\textrm{em}}/kT_{\textrm{em})}-1} d\nu_\textrm{em}.
\end{equation}

Here $h$ and $k$ are the Planck and Boltzmann constants, respectively, and $\beta$ is the dust emissivity index that we fix to a value of 2.\\
This assumption is reasonable when using temperatures derived from optically-thin models \citep[e.g.,][]{Casey_2012}, but care must be taken when interpreting these temperatures near the peak of the SED for DSFGs with ULIRG-like luminosities, where the dust may not be optically thin and opacity effects could become significant.

We know that the observed frequency $\nu_\textrm{obs}$ and the observed dust temperature temperature $T_\textrm{obs}$ are related to the emitted frequency $\nu_\textrm{em}$ and the emitted dust temperature $T_\textrm{em}$ by
\begin{equation}
    \nu_{\textrm{em}} =\nu_\textrm{obs} (1+z), 
\end{equation}
\begin{equation}
\label{for_jacobian1}
    T_{\textrm{em}} = T_\textrm{obs} (1+z),
\end{equation}
% We cant to see how much this SED has shifted to the observed frame which is parameterized in terms of $A_o, T_o$, and$ \nu_{o}$, Thus rewriting this in observed frame using $d\nu_e = (1+z) d\nu_o$ yields:
%  $$ S_{\nu_o} (\nu_{o}, T_{e}, A_{e}) d\nu_o= A_{e}.\frac{2h}{c^2}\cdot\frac{(1+z)\nu_{e}^{3+\beta}}{exp(\frac{(1+z)h\nu_{o}}{kT_{e}})-1}\cdot(1+z)d\nu_o$$
hence we can rewrite Eq.~\ref{emitted_modified_bb} in terms of observed-frame quantities as

 \begin{equation}
 \label{sed_to_fit}
       S_{\nu_\textrm{obs}} ( T_{\textrm{obs}}, A_{\textrm{obs}}) d\nu_\textrm{obs}= A_{\textrm{obs}}\frac{\nu_{\textrm{obs}}^{3+\beta}}{\text{exp}(h\nu_{\textrm{obs}}/k T_{\textrm{obs}})-1} d\nu_\textrm{obs},
 \end{equation} 
where we have used
  \begin{equation}
  \label{amplitude_eq}
      A_\textrm{obs} =  A_{\textrm{em}}(1+z)^{4+\beta}.
  \end{equation}

Using this SED model, we can determine the optimal values of the two unknown observed-frame parameters, $A_\textrm{obs}$ and $T_\textrm{obs}$, by fitting the model to the three SPIRE flux density measurements at 250, 350 and $\SI{500}{\mu m}$. Here we fit for these parameters
% involves minimizing the likelihood function $p(A_\textrm{o}, T_\textrm{o}| S_{250}, S_{350}, S_{500})$, 
using a Markov chain Monte Carlo (MCMC), which often yields the best uncertainty estimates; however, a simpler minimizer method could also have been used to achieve similar results given the limited number of data points.

It is worth mentioning that when converting to the emitted frame, $T_{\textrm{em}}$ and $1+z$ are entirely degenerate, since $\nu_\textrm{obs}/T_{\textrm{obs}} = \nu_\textrm{em}/T_{\textrm{em}}$.
% the fit becomes entirely degenerate when considering the parameter combination $T_\textrm{e}/(1+z)$. 
Therefore, obtaining a redshift estimate necessitates knowledge of $T_\textrm{em}$, and the uncertainty in the redshift estimation is primarily driven by the broad intrinsic temperature distribution \citep[e.g.,][]{Wuyts2008}. Nevertheless, we can improve the accuracy of the redshift estimate by incorporating information from the luminosity, as we now show.

In general, a luminosity $L$ can be related to a flux (the flux density integrated over a given wavelength range) and luminosity distance $D_{\textrm{L}}$.  
% \RH{I got rid of the luminosity density ($L_{\nu}$) and replaced it with flux ($F$) as I think it is a bit easier to understand the math. I also deleted a line of math as I think it could be described in words. I also changed $L_{\rm tot}$ to $L_{\rm IR}$.}
% \begin{equation}
%     L = 4 \pi D_\textrm{L}(z)^2 F ,
% \end{equation}
 A common wavelength range comes from defining the `total' infrared luminosity as the integral from 8 to \SI{1000}{\micro\meter} in the rest frame, so that
 \begin{equation}
    L_\textrm{IR} = 4 \pi D_\textrm{L}(z)^2 
    \int_{1000}^{8} S_{\nu_\textrm{em}} d\nu_\textrm{em}.
\end{equation}
Note that `8' and `1000' should be understood to mean the frequencies associated with these wavelengths in $\mu$m.

Next, we have a constraint equation given by the observed correlation between the temperature and the total infrared luminosity of galaxies. Following studies of this correlation in nearby and distant infrared-luminous galaxies \citep[e.g.][]{Chapin_2009}, we model this as a power law (although the functional form could be more general):
\begin{equation}
\label{intro:power_law}
     L_{\textrm{IR}} = f(T_{\textrm{em}}) =  \alpha T_{\textrm{em}}^{\gamma} .
\end{equation}

Finally, since 
\begin{equation}
    F_{\textrm{IR}} = \frac{L_{\textrm{IR}}}{4\pi D_\textrm{L} (z)^2} =  \int_{1000}^{8} S_{\nu_\textrm{em}}( T_\textrm{em}, z)d\nu_\textrm{em} \ , 
\end{equation}
we can substitute Eqs.~\ref{emitted_modified_bb}, \ref{amplitude_eq}, and \ref{intro:power_law} to obtain
% \begin{equation}\label{for_jacobian2}
%  A_\textrm{obs} = \frac{1}{\int_{1000}^{8}\left(\frac{\nu_{\textrm{em}}}{1+z}\right)^{3+\beta} \left[\text{exp}\left(\frac{h\nu_\textrm{em}}{kT_\textrm{em}}\right)-1\right]^{-1} \frac{d\nu_\textrm{em}}{1+z}} \frac{\alpha T_{\textrm{em}}^{\gamma}}{4 \pi D_\textrm{L}(z)^2}. 
% \end{equation}
\begin{multline}
\label{for_jacobian2}
     A_\textrm{obs} = \left (\frac{\alpha T_{\textrm{em}}^{\gamma}}{4 \pi D_\textrm{L}(z)^2}\ \right) \Big/ \\
     \left (\int_{1000}^{8}\left(\frac{\nu_{\textrm{em}}}{1+z}\right)^{3+\beta} \left[\text{exp}\left(\frac{h\nu_\textrm{em}}{kT_\textrm{em}}\right)-1\right]^{-1} \frac{d\nu_\textrm{em}}{1+z} \right). 
\end{multline}
%then
%\begin{equation}
%    S_{\textrm{rest}} =  \int_{1000}^{8} A_{\textrm{o}}\frac{2h}{c^2}\left(\frac{1}{1+z}\right)^{3+\beta} \frac{\nu_\textrm{e}^{3+\beta}}{\text{\text{exp}}\left(\frac{h\nu_\textrm{e}}{kT_\textrm{e}}\right)-1} \frac{d\nu_\textrm{e}}{1+z}
%\end{equation}
%and we obtain
%\begin{equation}
% A_\textrm{o} = \frac{1}{\int_{1000}^{8}\frac{2h}{c^2}\left(\frac{\nu_{\textrm{e}}}{1+z}\right)^{3+\beta} \left[\text{exp}\left(\frac{h\nu_\textrm{e}}{kT_\textrm{e}}\right)-1\right]^{-1} \frac{d\nu_\textrm{e}}{1+z}} \frac{L_\textrm{IR}}{4 \pi D_\textrm{L}(z)^2}. 
%\end{equation}

% \begin{equation}
% T_o = \frac{T_e}{(1+Z)}
% \end{equation}
The goal now is to go from the space of observed amplitude ($A_\textrm{obs}$) and observed temperature ($T_\textrm{obs}$), to the space of redshift ($z$) and emitted temperature ($T_\textrm{em}$). We can do this using a Jacobian transformation:
\begin{align}
\label{intro:transformation}
& p(z, T_\textrm{em})\, dz\, dT_\textrm{em} = \nonumber \\
&p[A_\textrm{obs}(z, T_\textrm{em}), T_\textrm{obs}(z, T_\textrm{em})] \cdot \textrm{det}
\begin{vmatrix}
\frac{\partial A_\textrm{obs}}{\partial z} & \frac{\partial A_\textrm{obs}}{\partial T_\textrm{em}} \\
\frac{\partial T_\textrm{obs}}{\partial z} & \frac{\partial T_\textrm{obs}}{\partial T_\textrm{em}}
\end{vmatrix} dz\, dT_\textrm{em}.
\end{align}
This can then be marginalized over $T_\textrm{em}$ to produce a one-dimensional probability density function (PDF) for redshift, $p(z)$. Let us calculate each cell of this Jacobian explicitly using Eqs.~\ref{for_jacobian1} and \ref{for_jacobian2}. Firstly,
\begin{multline}
\frac{\partial A_\textrm{obs} (z, T_{\textrm{em}})}{\partial z} = \\
\frac{\partial}{\partial z} \left[\frac{1}{\int_{1000}^{8} S_{\nu_{\textrm{em}}} (T_\textrm{em},z)\,d\nu_\textrm{em}} \right]\frac{f(T_{\textrm{em}})}{4\pi D_\textrm{L}(z)^2} \\
+ \frac{1}{\int_{1000}^{8} S_{\nu_{\textrm{em}}} (T_\textrm{em},z)\,d\nu_\textrm{em}} \frac{\partial}{\partial z}\left[ \frac{f(T_{\textrm{em}})}{4\pi D_\textrm{L}(z)^2}\right] \\ 
= -\frac{\left(\int_{1000}^{8}\frac{\partial}{\partial z} S_{\nu_{\textrm{em}}} (T_\textrm{em},z)\,d\nu_\textrm{em} \right)}{\left(\int_{1000}^{8} S_{\nu_{\textrm{em}}} (T_\textrm{em},z)\,d\nu_\textrm{em} \right)^2} \frac{f(T_{\textrm{em}})}{4\pi D_\textrm{L}(z)^2} \\
+ \frac{1}{\int_{1000}^{8} S_{\nu_{\textrm{em}}} (T_\textrm{em},z)\,d\nu_\textrm{em}} \left[\frac{-2 f(T_{\textrm{em}})}{4\pi D_\textrm{L}(z)^2} \frac{\partial D_\textrm{L}(z)}{\partial z} \right],
\end{multline}
where $\partial D_\textrm{L}(z)/\partial z $ can be performed numerically assuming \citet{Planck2018results} cosmological parameters.
The partial derivative is given by 
% \RH{Technically this next equation should have a factor of $2h/c^2$, although I suggest dropping that constant from all the equations entirely!}
\begin{multline}
    \frac{\partial}{\partial z} [S_{\nu_{\textrm{em}}} ( T_\textrm{em}, z) d\nu_\textrm{em}] =  \\
     -(4 + \beta) \left(\frac{1}{1+z}\right)^{5+\beta} \frac{\nu_{\textrm{em}}^{(3+\beta)}}{\text{exp}\left(\frac{h\nu_{\textrm{em}}}{kT_{\textrm{em}}}\right)-1}.
\end{multline}
The next term in the Jacobian is
\begin{multline}
\frac{\partial A_\textrm{obs}(z, T_{\textrm{em}})}{\partial T_{\textrm{em}}} = \\
\frac{\partial}{\partial T_{\textrm{em}}} \left[\frac{1}{\int_{1000}^{8} S_{\nu_{\textrm{em}}} (T_\textrm{em},z)d\nu_\textrm{em}} \right]\frac{f(T_{\textrm{em}})}{4\pi D_\textrm{L}(z)^2}\\ + \frac{1}{\int_{1000}^{8} S_{\nu_{\textrm{em}}} (T_\textrm{em},z)d\nu_\textrm{em}} \frac{\partial}{\partial T_{\textrm{em}}}\left[ \frac{f(T_{\textrm{em}})}{4\pi D_\textrm{L}(z)^2}\right] \\
 = -\frac{\left(\int_{1000}^{8}\frac{\partial}{\partial T_{\textrm{e}}} S_{\nu_{\textrm{em}}} (T_\textrm{em},z)d\nu_\textrm{em} \right)}{\left(\int_{1000}^{8} S_{\nu_{\textrm{em}}} (T_\textrm{em},z)d\nu_\textrm{em} \right)^2}\frac{f(T_{\textrm{em}})}{4\pi D_\textrm{L}(z)^2}\\
+ \frac{1}{\int_{1000}^{8} S_{\nu_{\textrm{em}}} (T_\textrm{em},z)d\nu_\textrm{em}} \left[\frac{1}{4\pi D_\textrm{L}(z)^2} \frac{\partial f(T_{\textrm{em}})}{\partial T_{\textrm{em}}} \right],
\end{multline}
and the remaining partial derivatives are 
% \RH{This would also need a $2h/c^2$}
\begin{multline}
    \frac{\partial}{\partial T_{\textrm{em}}} S_{\nu_{\textrm{em}}} ( T_\textrm{em}, z) = \\ \left(\frac{1}{1+z}\right)^{4+\beta} \frac{\nu_{\textrm{em}}^{(3+\beta)}}{\left[\text{exp}\left(\frac{h\nu_{\textrm{em}}}{kT_{\textrm{em}}}\right)-1\right]^2} \left( \frac{h\nu_{\textrm{em}}}{kT_{\textrm{em}}^2} \right) \text{exp}\left(\frac{h\nu_{\textrm{em}}}{kT_{\textrm{em}}}\right),
\end{multline}
 \begin{equation}
     \frac{\partial f(T_{\textrm{em}})}{\partial T_{\textrm{em}}} = \alpha \gamma T_{\textrm{em}}^{\gamma -1} .
\end{equation}

The last two terms in the Jacobian are derivatives with respect to the observed temperature, and are given by
\begin{equation}
  \frac{\partial T_\textrm{obs}}{\partial z} = -\frac{T_\textrm{em}}{(1+z)^2} 
\end{equation}
and 
\begin{equation}
    \frac{\partial T_\textrm{obs}}{\partial T_\textrm{em}} = \frac{1}{1+z} .
\end{equation}
By explicitly computing the Jacobian, multiplying by the two-dimensional function $p[A_\textrm{obs}(z, T_\textrm{em}), T_\textrm{obs}(\textrm{z}, T_\textrm{em})]$, and marginalizing over the range of $T_\textrm{em}$, we obtain
\begin{multline}
p(z) =
\int p[A_\textrm{obs}(z, T_\textrm{em}), T_\textrm{obs}(z, T_\textrm{em})\times\\ \left[ \frac{\partial A_\textrm{obs}}{\partial z} \frac{\partial T_\textrm{obs}}{\partial T_\textrm{em}} - \frac{\partial A_\textrm{obs}}{\partial T_\textrm{em}}\frac{\partial T_\textrm{obs}}{\partial z}\right] dT_\textrm{em} .
\end{multline}

The luminosity-temperature relation is of course not exact, therefore it is important to also marginalize over a width representing the intrinsic variance of galaxy infrared luminosities for a given temperature.
% include a variable that represents a correction factor that quantifies the deviation of an individual galaxy from the best-fit of the $L$--$T$ relation. 
%Empirically, the scatter is around 0.1\,dex in luminosity, as outlined by \cite{Chapin_2009}
To incorporate the scatter in the relation, during the MCMC process we also draw a Gaussian random variable with a mean centered at $\alpha$ and a standard deviation of $0.22 \alpha$ (as explained in the next subsection and shown in Fig.~\ref{fig:LT2}, also see subsection 3.2 of \citealt{Chapin_2009}); in other words we include a deviation in $\log L_{\rm IR}$
about the mean relation.  We then recalculate the Jacobian for each new $\alpha$, thus marginalizing over the intrinsic dispersion in the luminosity-temperature relation, which effectively increases the width of the redshift probability distribution. It is worth pointing out that the slope $\gamma$ could also be uncertain and potentially correlated with $\alpha$, but we are ignoring this for now.
% This approach allows us to capture the influence of the scatter in the $L$--$T$ relation and properly account for the variations observed in the data.

A key difference between this work and \citet{Aretxaga2003} is that we do not include any priors based on the expected redshift distribution for a given survey catalog. This is achieved in \citet{Aretxaga2003} by incorporating a range of SED templates and an empirical model for the evolution of the luminosity function in the context of the expanding Universe. Simulated galaxies are randomly drawn from this model with the appropriate survey parameters (such as flux limits at particular wavelengths) for comparison with observations of real galaxies selected in the same way. The posterior redshift PDF of a real galaxy is then derived from the redshifts of simulated galaxies with similar observed flux densities and colors. Redshifts with few (or no) galaxies in the simulated catalog will thus be heavily down-weighted. Such priors (if well-characterized) have the potential to yield estimates with narrower distributions (particularly for submillimeter surveys considered in isolation). Instead, our method simply assigns a greater weight to redshift estimates by applying a Gaussian prior on the temperature given the luminosity, ensuring that the implied rest-frame luminosity and temperature are physically consistent. Although this approach uses less information (and may therefore yield broader distributions), we believe it provides sufficient constraining power to reject catastrophic failures from otherwise optical and NIR-based photometric redshifts, as we will show.  In the future, as we obtain more information on the evolving luminosity function, we could add redshift priors.

To summarize, we used the three SPIRE photometric points to estimate two observed parameters, namely the peak frequency and amplitude, and then relate them to intrinsic quantities $T_\textrm{em}$ and $z$ using the luminosity-temperature relation and include an additional marginalization over the amplitude of this relation.

\begin{figure}[htbp!]
\centering
\includegraphics[scale = 0.384]{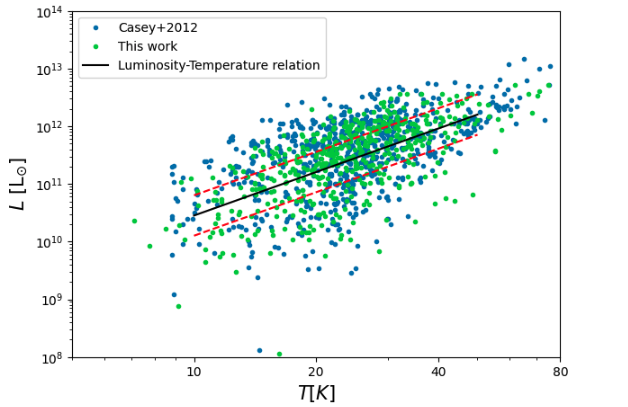}
\caption{FIR luminosity versus dust temperature. The blue points are taken directly from \citet{Casey_2012}, while the green points are our calculated luminosities, obtained by fitting the same SPIRE flux densities to Eq.~\ref{emitted_modified_bb} and integrating. The best-fit power-law provides our luminosity-temperature relation, shown in black. The dashed red lines represent the estimated intrinsic scatter (after removing the scatter due to statistical errors) and are given by ${\pm}0.22\alpha$.}
\label{fig:LT2} % Assign a unique label name
\end{figure}

\subsection{The galaxy luminosity-temperature relation}
To find the luminosity-temperature constraint parameters $\alpha$ and $\gamma$, we need to know the spectroscopic redshifts and emitted temperatures of a sample of galaxies. This correlation has been investigated by \citet{Chapin_2009} and \citet{Casey_2012}; the former used a sample of {\it IRAS\/}-selected low-$z$ galaxies, and the latter used a sample of {\it Herschel\/}-selected galaxies out to $z\,{=}\,2$. Here we primarily make use of the sample from \citet{Casey_2012}; Fig.~\ref{fig:LT2} shows the resulting luminosity-temperature correlation of their sample (using both the luminosity and temperature values provided in the paper and our own best-fit values using Eqs.~\ref{for_jacobian1} and \ref{sed_to_fit}). We find a best-fit amplitude of $\alpha\,{=}\, 10^{7.95}$ and a best-fit slope of $\gamma\,{=}\, 2.50$, and we confirm that our luminosity-temperature relation agrees with the results from the low-$z$ sample of \citet{Chapin_2009} by comparing with their Figure~4 (after implementing the necessary transformations, such as converting far-IR color to temperature).

In addition, we use the sample of SPIRE-selected galaxies from \citet{Casey_2012} to estimate the intrinsic scatter in the relation by symmetrically adjusting $\alpha$ about its best-fit value until $68\%$ of the galaxies fall within the lower and upper bounds, finding a range of $\pm 0.35$\,dex. This value is the uncertainty due to both intrinsic scatter and statistical noise in the SPIRE data, and for our photo-\textit{z} marginalizations we only want to take into account the intrinsic scatter. The median uncertainty in the $L_{\textrm{IR}}$ measurements of the sample was found to be 0.27\,dex, thus the intrinsic scatter is $\sqrt{0.35^2\,{-}\,0.27^2}\,{=}\,0.22$\,dex.

\begin{figure*}
\centering
\includegraphics[scale = 0.76]{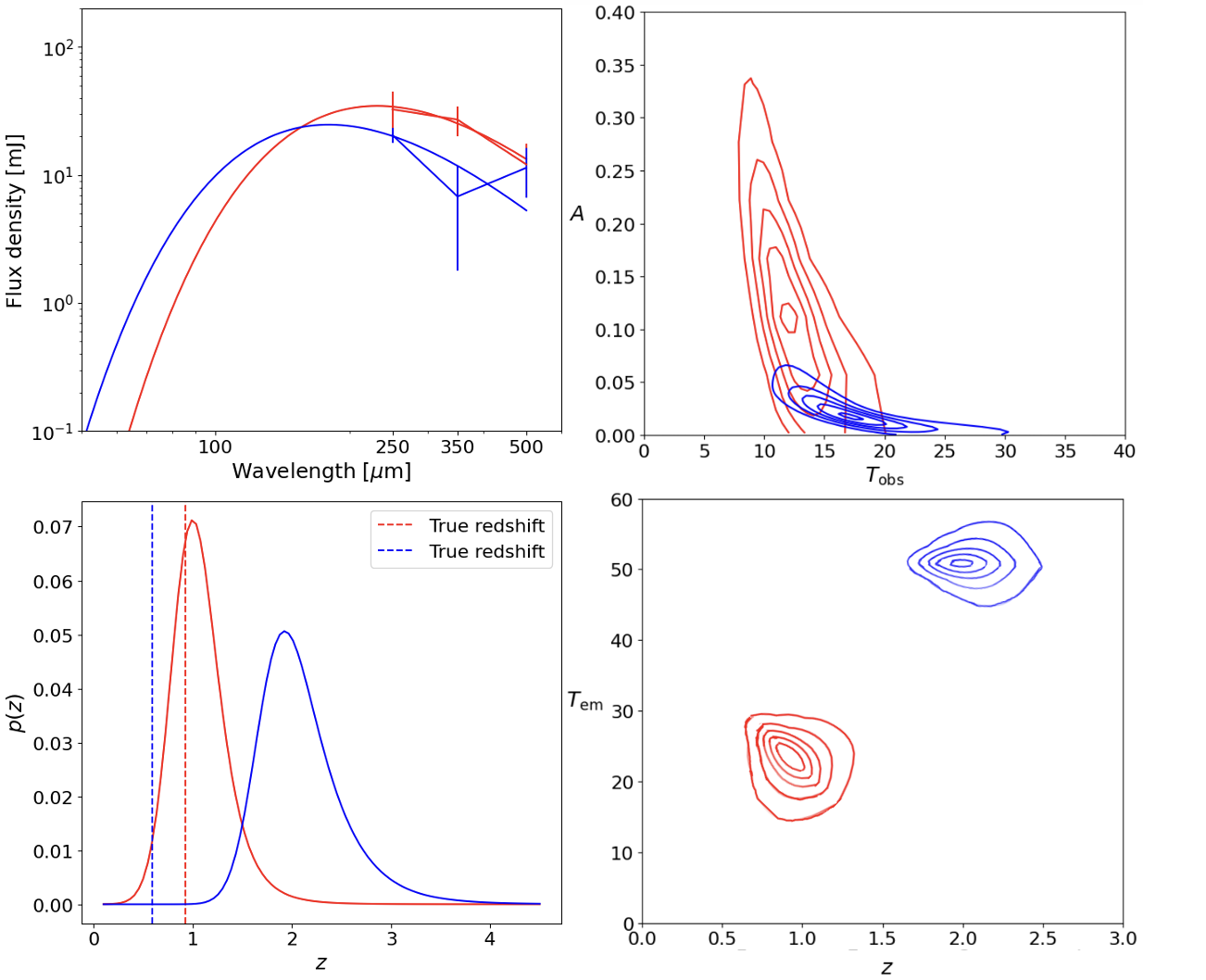}
\caption{Comparison between two galaxies, one with three submm flux density measurements that correctly identifies the peak of the modified blackbody function and leads to a reliable redshift estimate (red color), while the three fluxes of the galaxy in blue does not identify the peak of the function, which leads to an inaccurate estimate of the redshift. The contour plot of $A_\textrm{obs}$ and $T_\textrm{obs}$ (top right) are transformed using Eq.~\ref{intro:transformation} to obtain the corresponding contour plot for $T_\textrm{em}$ and $z$ (bottom right). The resulting redshift PDF are shown with solid red and blue lines while the actual redshift values are shown by dashed lines.} 

\label{fig:MCMC_good}
\end{figure*}

It is important to point out that biases in the sample used to calibrate the luminosity-temperature relation could result in inaccurate photo-\textit{z} estimation. For instance, the redshifts in the \citet{Casey_2012} sample are limited to values below 2, with the majority concentrated between 0.6 and 0.9. 
% it is important to note that all galaxies in the catalog have redshifts below 2, with the majority falling between 0.6 and 0.9. Consequently, when employing the MCMC, we are predisposed to predict redshifts within this range given three new \textit{SPIRE} flux densities. This observation gains further validation through simulated data, which shows diminishing reliability for galaxies at higher redshifts (e.g., $z > 2$) because the $L$--$T$ fit parameters are based on lower redshift galaxies. 
Consequently, when using this method to predict redshifts based on three SPIRE flux densities, predictions tend to cluster within this range. This observation is reinforced using simulated data, revealing a decrease in prediction reliability for galaxies at higher redshifts (in this case $z > 2$). This arises from the fact that the luminosity-temperature fit parameters are derived primarily from galaxies at lower redshifts. The method works well when applied to galaxies with similar redshifts as those used to establish the correlation parameters, and the range can be extended with future data sets. What we describe here is only an example.%  By incorporating samples of higher-redshift galaxies from future surveys we can look for redshift dependence in the $L$--$T$ relation, and thus make it more applicable to galaxies at both low and high redshifts.

\subsection{Submm/mm SED fitting}

\begin{figure}[htbp!]
\centering
\includegraphics[scale = 0.37]{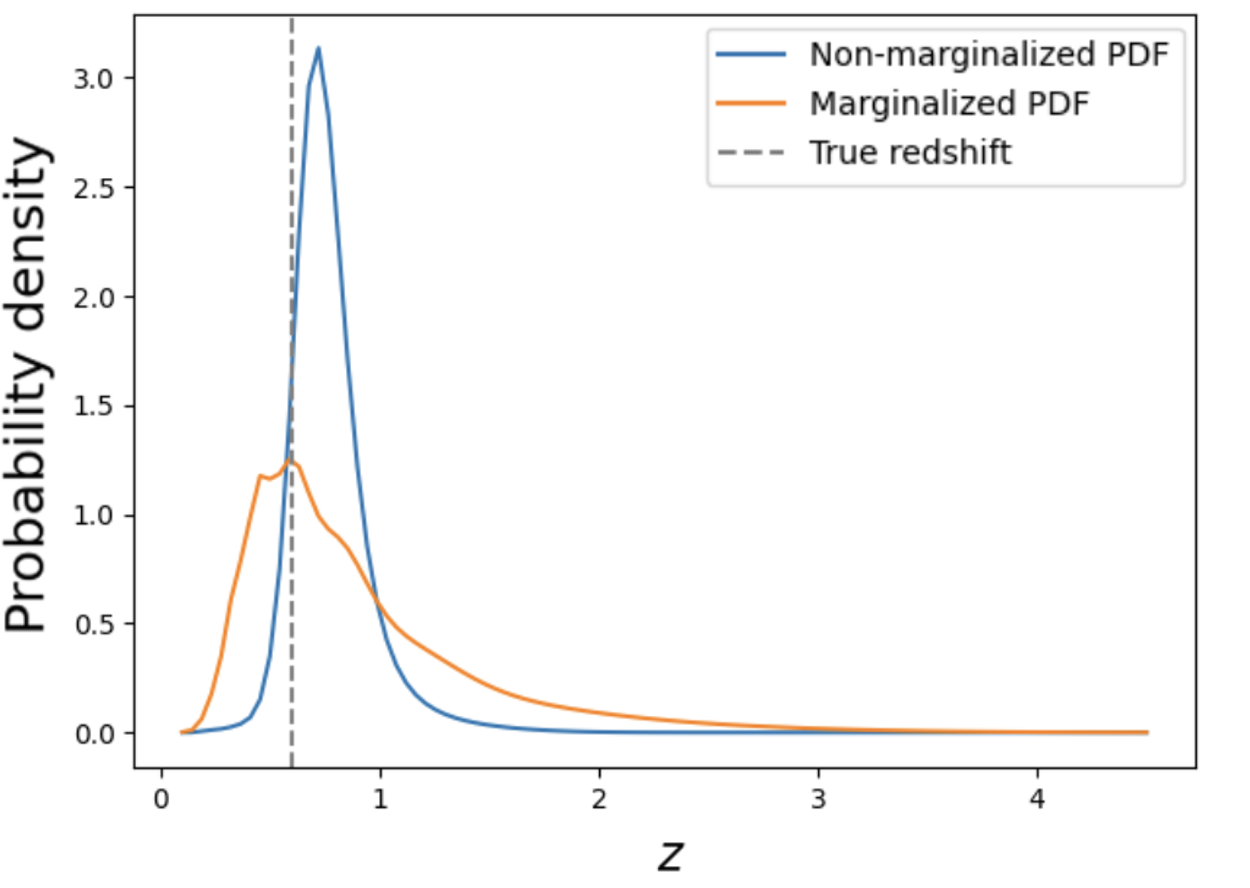}
\caption{An example showing the effect of marginalizing over $\alpha$ (see Eq.~\ref{intro:power_law}). The blue PDF is generated assuming that the luminosity-temperature relation is exact, while the orange PDF is computed by marginalizing over the amplitude of the luminosity-temperature relation.}
\label{fig:scatter}
\end{figure}

We fit the observed quantities with a standard MCMC method with no priors on the parameters using Python's \texttt{lmfit}, \citet{lmfit}, and minimize using the `emcee' method. The likelihood function is therefore simply given by $\exp(-\chi^2/2)$. After transforming from the 2D parameter space $(A_\textrm{obs}, T_\textrm{obs})$ to $(T_\textrm{em}, z)$, we obtain the PDF $p(z)$ by marginalizing over $T_\textrm{em}$.

Given the limited amount of information provided by only three submm data points, we can make accurate inferences about the shape of the SED only if these points provide meaningful information about the curvature of the SED. However, when all three flux density points are situated deep into the Rayleigh-Jeans or Wien side of the modified blackbody peak, or if blending with nearby galaxies contaminates the photometry, it is difficult to determine the peak position and amplitude. 
% This is due to a degeneracy in the amplitude and temperature parameters, so we cannot discern whether the observed amplitude should have been smaller or if the observed temperature should have been larger. 
Figures~\ref{fig:MCMC_good} provide an illustration of two scenarios: one with three flux densities containing meaningful curvature information, which yields a good constraint; and the other where the flux densities have been contaminated by source blending.
% degenerate outcomes on the Rayleigh–Jeans side. 
In cases where the modified blackbody model provides a poor fit (which we take to be $\chi^2\,{>}\,3 $ ), we disregard the FIR redshift PDF and only use the results from optical photometry. As discussed above, we also include an additional marginalization over $\alpha$ due to the intrinsic scatter in the luminosity-temperature relation, which broadens the probability distribution; and example of this is shown in Fig.~\ref{fig:scatter}. 
% \RH{Explain the criteria for deciding how a fit is a bad fit, and explain what you do to deal with these situations.}

\section{Combining FIR/submm and optical/NIR photometric redshifts}\label{sec:optical}
% Photometry in near-infrared and optical bands are considerably more challenging to model. Unlike in the far-infrared, where a simple SED can be assumed, galaxy SEDs in the optical bands are much more complicated, requiring more free parameters to fit. However, there are methods available to address this issue.

Although usually much more accurate than with submm data alone, Photometric redshift estimation using NIR and optical photometry typically provides greater constraining power on the redshift probability distribution $P(z)$ than even the best submm data, since high-quality OIR data are generally easier to obtain. However, this approach remains considerably more complicated due to the variety of SEDs involved, requiring additional parameters. The two standard approaches are the empirical method, which uses a training set to fit for the parameters of a simple function, and the SED template-fitting technique, which uses a library of possible template SEDs \citep[for example from][]{bruzual2003}.

To achieve our objective of combining optical photometric redshifts with submm photometric redshifts, we employ a code that uses the SED template-fitting approach. Specifically, we use the {\tt Easy Accurate zphoto from Yale} \citep[\texttt{EAzY};][]{EAZY} code, which computes linear combinations of SED templates in the optical-to-NIR over the redshift range of $0$--$4$. Importantly, the output of \texttt{EAzY} is a redshift PDF, which we can multiply by our FIR/submm redshift PDFs to obtain a final redshift estimate as the two distributions are completely independent.
% \texttt{EAzY} employs a flexible approach by including complex star-formation histories and fitting non-negative linear combinations of templates. At each redshift, the algorithm evaluates the likelihood function by determining the single best $\chi^2$ value among the various template combinations. 

We ran \texttt{EAzY} using the default parameter file, which is designed to provide an optimized template set, template error function, and priors for most applications.
While the majority of galaxy redshift estimates produced by \texttt{EAzY} will exhibit a single, well-defined peak, certain cases can yield multiple peaks due to degeneracies in the model parameters. This is can be particularly pronounced when dealing with specific subsets of galaxies, such as DSFGs or active galactic nuclei (AGN). Notably, galaxies with relatively featureless SEDs can equally match observations at both $z \sim 1$ and $z \sim 3$ \citep[e.g.,][] {Roseboom_2011}, and  
% This comes from the templates' inability to discern subtle variations in SEDs caused by features lying beyond the Balmer and Lyman breaks, as detailed in \citet{EAZY}.
an illustrative example of this is presented in Fig.~\ref{fig:EAzY1}. 

Finally, Fig.~\ref{fig:example_prob} shows an example of combining (i.e. multiplying) the results from both FIR/submm SED fitting and optical/NIR SED fitting. This particular example is a galaxy bright in the SPIRE wavelengths and faint in the optical, and we see that the photometric redshift estimate is improved by including longer-wavelength information.
%This challenge becomes particularly pronounced when dealing with specific subsets of galaxies, such as DSFGs and AGN. 
%\setstcolor{red}

\begin{figure}[htbp!]
\centering
\includegraphics[scale = 0.415]{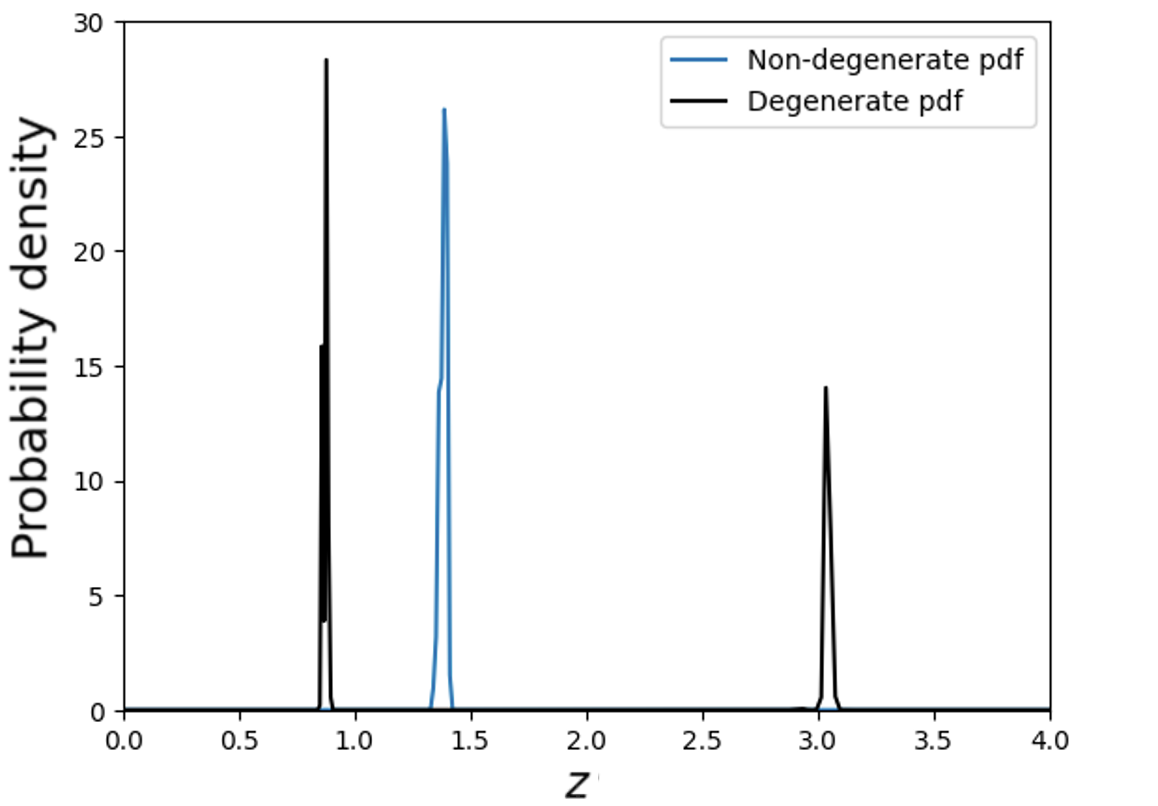}
\caption{Example redshift PDFs generated using \texttt{EAzY}. The black PDF illustrates a case of a galaxy with degenerate redshift values, whereas the blue PDF shows a galaxy with a distinct, non-degenerate (single) peak. We expect many DSFGs to fall in the former case, and we can break the degeneracy with information from FIR/submm wavelengths.}
\label{fig:EAzY1}
\end{figure}

\begin{figure}[htbp!]
\centering
\includegraphics[scale = 0.415]{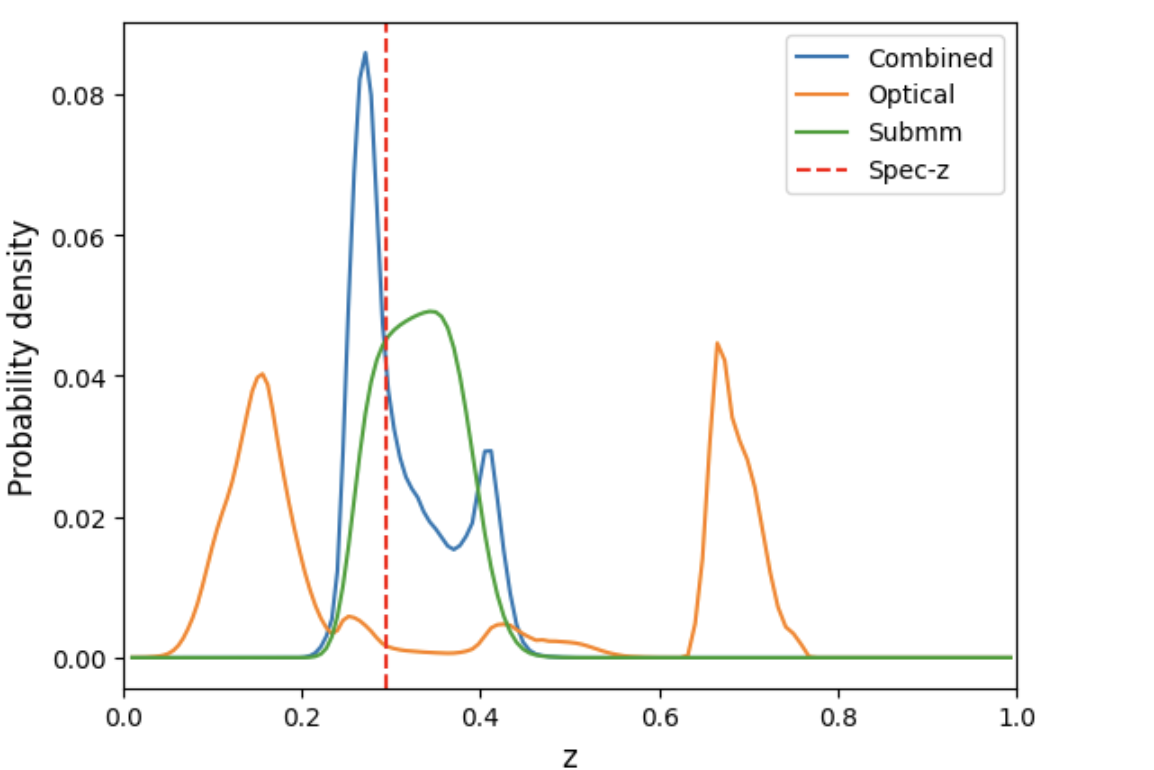}
\caption{Example redshift PDFs from \texttt{EAzY}, FIR photometry and the their combination (i.e. the product of the two probabilities), compared to the spectroscopic redshift.}
\label{fig:example_prob}
\end{figure}

\section{Applications to real galaxy catalogs}\label{sec:combining}
\subsection{Test catalogs}
To test our proposed method for combining optical/NIR photometry with FIR/submm photometry, we use the `super-deblended' galaxy sample in the COSMOS field \citep{Jin_2018}. 
This catalog uses optically-selected galaxies as priors to fit PSFs to {\it Herschel\/}-SPIRE data, thus providing the necessary optical-through-FIR photometry needed to test our proposed photometric-estimation method. The final deblendeded catalog includes 11{,}220 galaxies over the 100--1200\,$\mu$m range, extending to redshifts above 4. However, it is worth noting that SPIRE uncertainties in this catalog may be underestimated, since the accuracy of the deblending could be influenced by source confusion. To account for this, we applied a floor to the SPIRE uncertainties consistent with the confusion limits, since this provides a more realistic assessment of the data quality. From this catalog, we selected about 5\% (600) of the brighter galaxies ($K<22.5$) that have reliable optical/NIR photometry and spectroscopic redshift from COSMOS 2015 \citep
{Laigle2016}.

This `super-deblended' catalog is purely optically-selected. We also test our method using a purely FIR-selected sample of galaxies with redshifts between 0 and 1 from the Herschel Astrophysical Terahertz Large Area Survey (H-ATLAS), extensively detailed in \citet{HATLAS1} and \citet{HATLAS2}. This survey spans three fields covering a total area of 161.6\,deg$^2$ along the celestial equator with each field spanning approximately 54\,deg$^2$, and contains overlapping optical/NIR imaging.
This catalog, which is a subset of the gamma field of H-ATLAS, contains 120{,}230 sources selected
%with noise levels measuring 7.4, 9.4, and 10.2\,mJy
at 250, 350, and $\SI{500}{\mu m}$, along with photometry for their optical and NIR counterparts. After filtering for all the galaxies with \st{reliable} spectroscopic redshift, as well as optical and submm photometric bands, we selected a random sample of 250 of the brighter galaxies for this analysis.

\subsection{Catastrophic outlier metric}
We define the catastrophic outlier fraction, $\eta$, to be the fraction of sources with an unexpectedly large difference between the photometric and spectroscopic redshift. This is explicitly given by
\begin{equation}\label{outlaw}
    \delta z> 3\sigma,
\end{equation}
where
\begin{equation}\label{outlier}
\delta z = \frac{\left| z_{\text{photo}}-z_{\text{spec}}   \right|}{1 + z_{\text{spec}}}
\end{equation}
and
\begin{equation}\label{sigma_def}
    \sigma_{\text{\textbf{NMAD}}} = 1.48 \times \text{median} \left( \left| \frac{\Delta z - \text{median}(\Delta z)}{1 + z_{\text{spec}}} \right| \right), 
\end{equation}
with $\Delta z = z_{\text{photo}} - z_{\text{spec}}.$ We have chosen this definition of $\sigma$ as it is less sensitive to outliers than the usual definition of the standard deviation \citep{EAZY}.
%We also let $N\,{=}\,3$, but this is an arbitrary choice depending on the ideal threshold for a study.

\subsection{Results}
We want to demonstrate that our method improves photo-$z$ estimates for existing samples of galaxies.  However, a challenge is that we can only test how well the estimates improve photo-$z$s if we have spec-$z$s for the sources.  Good spectroscopic redshifts are still scarce for submm-selected galaxies, while optically-selected catalogs with good spectroscopy often have such good photo-$z$s that the submm data hardly help.  We therefore start with an optically-selected sample, and artificially degrade it to represent a catalog with only moderately-constraining optical/NIR photometry.

For this test, we use a subset of six \textit{Euclid}-like photometric bands (B, R, Ks, J, H, and Y)
% \RH{explicitly write out the six bands you used }
from the `super-deblended' catalog, and artificially increase the noise by a factor of 3 in order to simulate the type of data expected from future wide field surveys. No SED interpolation is performed to generate Euclid-like photometry; rather, bands that are closest to the Euclid filters are used.
The results are shown in Fig.~\ref{fig:comparison_same}; we see that while the fraction of catastrophic outliers is already low for these optically-bright galaxies, using submm photometry is still able to reduce the fraction further. 

We also tested this result by using the original `super-deblended' catalog (without removing bands or changing the noise). However, including submm bands did not provide a significant improvement, since the photometric redshifts of these optically-bright galaxies were already very closely aligned with the spectroscopic redshifts.

\begin{figure}[htbp!]
\centering
\includegraphics[scale = 0.40
]
{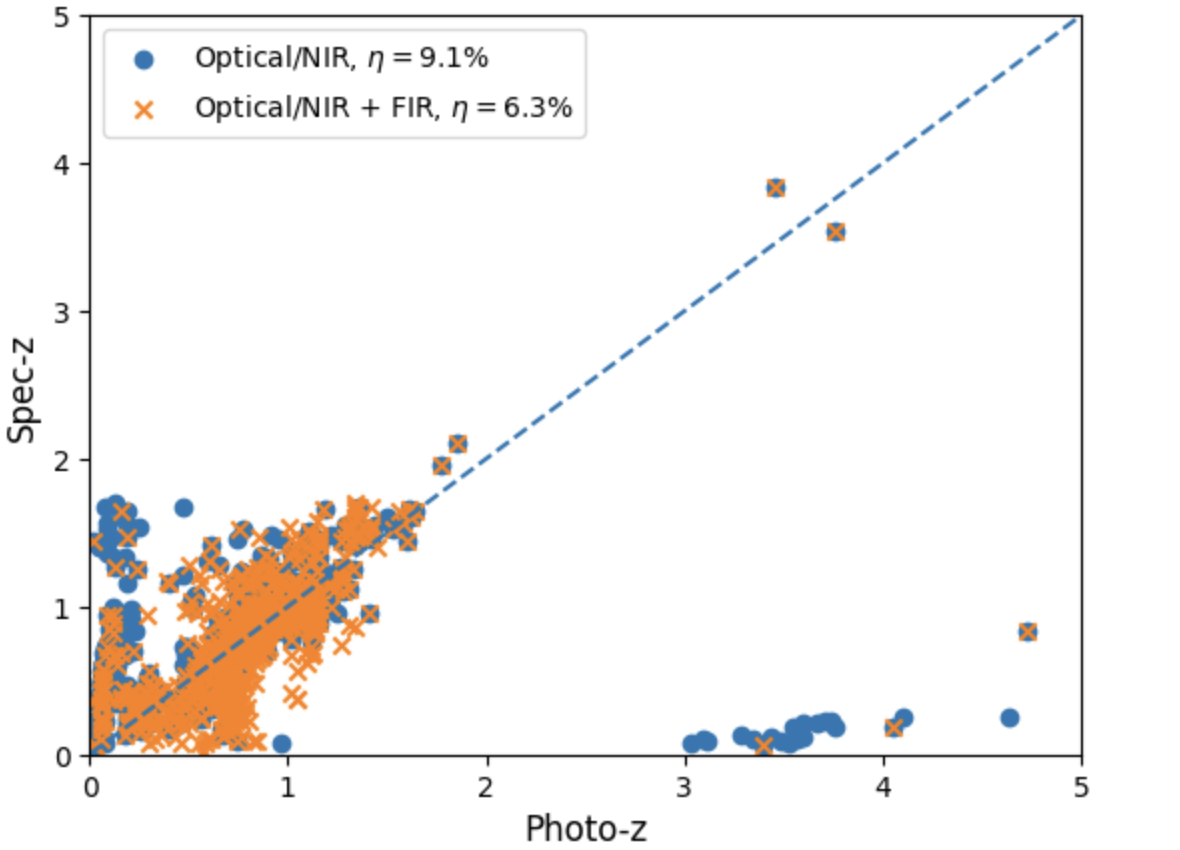}
\caption{\textbf{COSMOS OIR} Spec-\textit{z}s versus photo-\textit{z}s for a sample of 630 `super-deblended' galaxies in the COSMOS field from \citet{Jin_2018}, artificially degraded by restricting the optical/NIR photometry to six bands and increasing the noise by a factor of 3. Photometric redshifts using optical and NIR data with \texttt{EAzY} alone are shown in blue, while photometric redshifts after combining with submm data are shown in orange. Here the fraction of outliers for the blue data is $9.1 \%$, while for the orange data is $6.3 \%$.}
\label{fig:comparison_same}
\end{figure}

%\subsubsection{FIR-selected catalog}

We next turn to galaxies selected by SPIRE. For this test we used the U, G, R, I, Ks, J, H, Y, and Z bands. The result of applying our algorithm to the H-ATLAS catalog is shown in Fig.~\ref{fig:final_z_z}. Without adjusting any of the optical/NIR photometry, we find that the number of catastrophic outliers reduces from 23 (9.2\%) to 8 (3.2\%), showing a significant improvement in photometric redshift estimation.

\begin{figure}[htpb!]\label{final_plot}
\centering
\includegraphics[scale = 0.43]{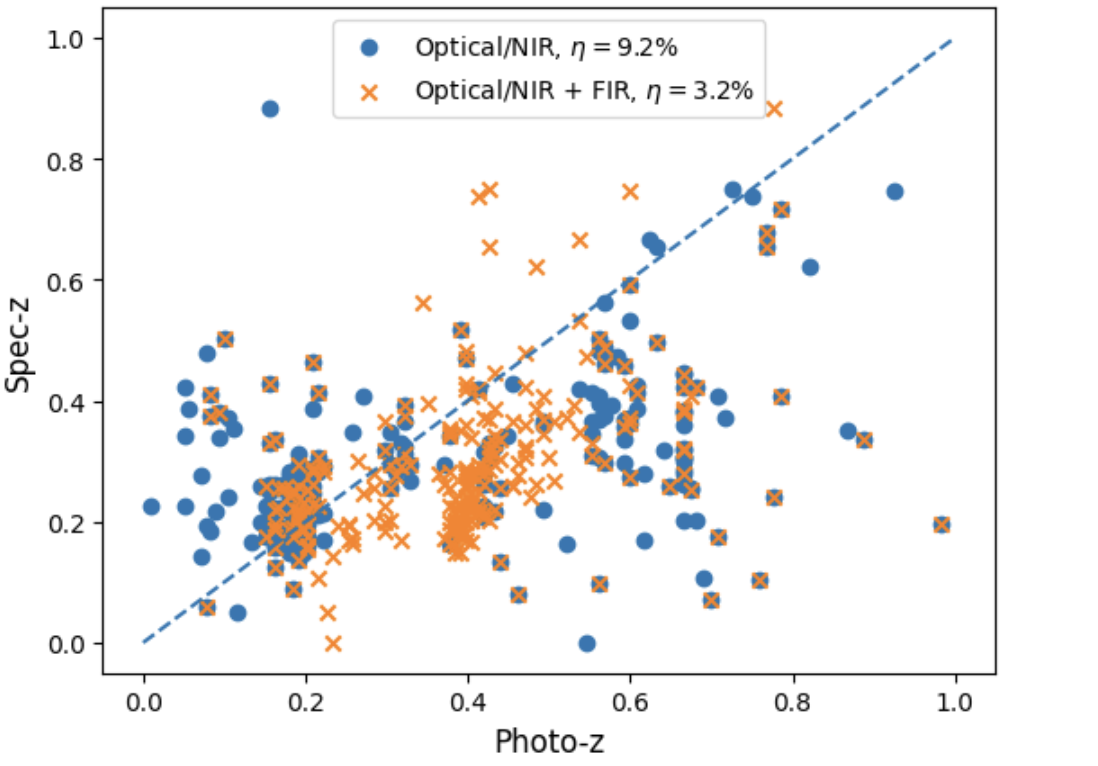}
\caption{Spec-\textit{z} versus photo-\textit{z} for a sample of 250 bright submm galaxies taken from the H-ATLAS catalog \citep{HATLAS1, HATLAS2}. The inclusion of submm bands provides a significant reduction in catastrophic outliers (defined by Eq. \eqref{outlier}), with three times fewer outliers compared to using only optical and NIR bands.}
%Here the fraction of outliers for the blue data is $9.2 \%$, while for the orange data is $3.2\%$.}
\label{fig:final_z_z}
\end{figure}

% \RH{More to say here?}
\subsection{Future Improvements}

While in this paper we used the specific example of \textit{Herschel}-SPIRE photometry, our approach is general, and adding more submm photometric points (e.g., $850\,\mu$m data from SCUBA-2) would further improve our method.  Additionally, as we learn more about the relevant galaxy populations, from \textit{Euclid} and Rubin surveys for example, we can continually update the $L$--$T$ relation, thus improving the submm photo-$z$ estimates.

Finally, we can also include a volume prior to better capture the galaxy population distribution, similar to the work done by \citet{Aretxaga2003}. Incorporating this prior here would not have made a significant difference because our test catalog of galaxies do not span a very large redshift range. However, with future data sets spanning a wider redshift range, the use of a volume prior will be much more important.

\section{Conclusions}\label{sec:conclusions}
The primary objective of this paper is to enhance photometric redshift estimates, specifically targeting the dustiest and most actively star-forming galaxies where conventional optical and NIR fits sometimes fall short. This method is particularly useful for wide-area surveys limited to optical data, such as those conducted with Rubin Observatory's Legacy Survey of Space and Time (LSST), where supplemental data in the far-infrared (FIR) from {\it Herschel} or future instruments like CCAT or TolTEC on the Large Millimeter Telescope (LMT) could dramatically improve constraints on these dusty sources. On the timescale of the next decade, these combined datasets will provide opportunities to refine redshift estimates for large galaxy samples in poorly explored redshift-luminosity regimes. By focusing on sources where optical and NIR fits are unreliable, our approach aligns with the overarching goal of improving photo-\textit{z} accuracy for all galaxies in wide-field surveys, enabling more robust constraints on galaxy evolution across cosmic time.

The largest photo-\textit{z} galaxy samples in the near future are going to come from the \textit{Euclid} and Rubin surveys. There is a strong emphasis on the use of galaxies for cosmological studies involving gravitational lensing and clustering, and hence galaxies with challenging photometric redshifts will tend to be excluded. However, for studies of galaxy evolution, we certainly do not want to throw away the DSFGs.
Looking ahead, the combination of Rubin photometry with FIR datasets already in hand or from future facilities like TolTEC and CCAT can help create more complete redshift catalogs for studies focused on dust production, extreme star-forming systems, and high-mass galaxies. While DSFGs constitute a small fraction of the galaxy population, they play a critical role in understanding the cosmic star-formation history and the evolution of massive systems. By applying our method to refine redshift estimates for these rare but significant galaxies, we aim to expand the utility of future extragalactic surveys for a broad range of galaxy evolution studies.

\begin{acknowledgments}
This work has been supported by the Natural Sciences and Research Council of Canada and the Canadian Space Agency.  We thank Caitlin Casey for helpful comments.
\end{acknowledgments}

%%%%%%%%%%%%%%%%%%%%%%%

\bibliography{ref}{}
\bibliographystyle{aasjournal}

\end{document}